**Exceptional points as lasing pre-thresholds**

*Alexander A. Zyablovsky,*[1,2] *Ilya V. Doronin,*[1,2] *Eugeny S. Andrianov,*[1,2] *Alexander A. Pukhov,*[1,2,3] *Yurii E. Lozovik,*[1,4] *Alexey P. Vinogradov,*[1,2,3] *and Alexander A. Lisyansky*[5,6]*

[1]Dukhov Research Institute of Automatics, 22 Sushchevskaya, Moscow, 127055, Russia
[2]Moscow Institute of Physics and Technology, 9 Institutskiy pereulok, Moscow, 141700, Russia
[3]Institute for Theoretical and Applied Electromagnetics, 13 Izhorskaya, Moscow, 125412, Russia
[4]Institute of Spectroscopy Russian Academy of Sciences, 5 Fizicheskaya, Troitsk, Moscow, 108840, Russia
[5]Department of Physics, Queens College of the City University of New York, Queens, New York, 11367, U.S.A.
[6]The Graduate Center of the City University of New York, New York, 10016, U.S.A.E-mail:

The genesis of lasing, as an evolution of the laser hybrid eigenstates comprised of electromagnetic modes and atomic polarization, is considered. It is shown that the start of coherent generation at the laser threshold is preceded by the formation of a special hybrid state at the *lasing pre-threshold*. This special state is characterized by an enhanced coupling among excited atoms and electromagnetic modes. This leads to an increase in the rate of stimulated emission in the special state and, ultimately, to lasing. At the lasing pre-threshold, the transformation of hybrid eigenstates has the features of an exceptional point (EP) observed in non-Hermitian systems. The special state is formed when eigenfrequencies of two hybrid states coalesce or come close to each other . Below the pre-threshold, lifetimes of all hybrid states grow with increasing pump rate. When the pump rate crosses the pre-threshold, resonance trapping occurs with the lifetime of the special state continuing to increase while the lifetimes of all other eigenstates begin to decrease. Consequently, the latter eigenstates do not participate in the lasing. Thus, above the pre-threshold, a laser transitions into the single-mode regime.

## 1. Introduction

Non-Hermitian systems possess many unusual features.[1-7] One of the most interesting is the presence of exceptional points (EPs) in the parametric space.[8,9] Namely, reaching an EP results in two or more system eigenstates becoming linearly dependent, as their eigenfrequencies coalesce.[8,9] When an EP is crossed, the properties of the system change. In the vicinity of the EP, a state transformation occurs which is a manifestation of the resonance trapping effect[10-14] (see also ref. [15] and references in therein). In this effect, the system eigenstates are separated into two types, when the spacing between two eigenstates becomes smaller than their linewidth. When the parameter characterizing system non-Hermiticity increases, the lifetimes of the first type of eigenstates increase, while the lifetimes of the second type of eigenstates decrease.[12,15]

Systems with EPs are employed in numerous applications. For example, they are used to enhance the sensitivity of laser gyroscopes[16] and sensors,[17-20] to select modes in multimode lasers,[21,22] and to achieve lasing without inversion.[23]

The most celebrated laser systems with EPs are lasers having parity–time (PT) symmetry,[21,22,24-30] which is achieved by balancing amplifying and absorbing regions. In these lasers, an EP is associated with breaking the PT-symmetry of eigenmodes.[31] It is manifested as a dramatic rebuilding of the electromagnetic (EM) fields in the eigenmodes. Some eigenmodes become predominantly localized within the amplifying medium, while the others are mainly localized within the absorbing medium. As a result, the former eigenmodes have a greater lifetime than the other eigenmodes. The mode transformation in an EP is used to achieve single-mode lasing in multimode systems.[21,22,27] In addition, this transformation can result in a non-monotonic dependence of the lasing threshold on the pump rate [28,29] or losses.[26]

In this paper, we demonstrate that EPs or at least signatures of EPs take place in the majority of conventional lasers, which have no particular symmetry. In contrast to PT-symmetric systems, in lasers, the EPs are not connected with a symmetry breaking. They appear because the eigenstates of a laser system are hybrid states of the EM field and the atomic polarization. Before an EP is reached, the contribution of the atoms' polarization into all eigenstates increases with an increase in the pump power. After an EP is crossed, the rate of this contribution into one of the eigenstates (the special state) increases. The increase of the atoms' polarization contribution into the special state results in the strengthening of the mode-active medium interaction and the enhancement of the light amplification in the special state. Consequently, the special state has the smallest lasing threshold, and a further increase in the pump rate results in lasing in this state. Therefore, the pump rate, at which the special state forms, can be considered as the pre-threshold for lasing. We show that lasing pre-thresholds take place in systems with and without cavities.

## 2. The model

To study a conventional multimode laser, we consider a 1D model of an ensemble of incoherently pumped atoms. We are interested in system behavior when the pump power is smaller than the lasing threshold. The light is generated in the volume occupied by an active medium and is radiated into the environment. We consider systems with and without resonators.

Following the standard procedure of second quantization,[32] we assume that the laser is placed in a one-dimensional box (waveguide) of the size $L_B$. This box plays the role of the Universe. For finite $L_B$, to avoid the impact of the waves reflected from the borders of the environment, one should work in the time-domain and consider times smaller than the round-trip time of light in the system. This requires rather cumbersome calculations. The way around this difficulty is to fill the box with a weakly absorbing medium and choose $L_B$ much larger than the decay length of the laser radiation in the medium. This procedure is equivalent to going over to the limit $L_B \to \infty$. In this case, we can work in the frequency domain.

To describe the evolution of a large (infinite) number of EM modes with pumped two-level atoms, we use the Maxwell-Bloch equations[33,34] for corresponding modes interacting with atoms.[35,36] Since we study an absorbing environment, the frequencies of the EM modes have negative imaginary parts. For simplicity, we assume that these parts are the same for all modes, $\text{Im}\,\omega = -\gamma_a$.

We consider systems having $N > 10^4$ two-level atoms with the transition frequency $\omega_{TLS}$. Using the Maxwell-Bloch equations is justified if the number of atoms is large ($N \gg 1$).[33,34,36] Then we obtain the following equations

$$\frac{da_n}{dt} = \left(-\frac{\gamma_a}{2} - i\Delta_n\right)a_n - i\sum_{m=1}^{N}\Omega_{nm}\sigma_m, \qquad (1)$$

$$\frac{d\sigma_m}{dt} = -\frac{\gamma_\sigma}{2}\sigma_m + i\sum_{n=-\infty}^{\infty}\Omega_{nm}a_n D_m, \qquad (2)$$

$$\frac{dD_m}{dt} = (\gamma_P - \gamma_D) - (\gamma_P + \gamma_D)D_m + 2i\sum_{n=-\infty}^{\infty}\Omega_{nm}\left(a_n^*\sigma_m - a_n\sigma_m^*\right), \qquad (3)$$

where $a_n$ is the amplitude of the EM field in the $n$th mode. The quantities $\sigma_m$ and $D_m$ are the polarization and the population inversion of the $m$th two-level atom, respectively, $\gamma_a$ is the relaxation rate of the EM modes of the empty waveguide, $\gamma_D$ and $\gamma_\sigma$ are the relaxation rates of the population inversion and polarization of the atoms, respectively, and $\gamma_P$ is the rate of incoherent pumping of the atoms. The quantity $\Delta_n = \omega_n - \omega_{TLS}$ is a detuning between the frequency of the $n$th mode, $\omega_n$ and the transition frequency of the atoms, $\omega_{TLS}$, and $N$ is the number of two-level atoms. The Rabi coupling constant of the $n$th mode with the $m$th atom is $\Omega_{nm}(x_m) = -\mathbf{E}_n(x_m) \cdot \mathbf{d}/\hbar$,[35,36] where $x_m$ is the position of the *m*th atom, $\mathbf{E}_n(x)$ is the electric field "per one photon" of the *n*th mode, and $\mathbf{d} = \langle e|e\mathbf{r}|g\rangle$ is the matrix element of the dipole moment of a two-level atom having ground $|g\rangle$ and excited $|e\rangle$ states. The EM field distribution in the *n*th mode, $\mathbf{E}_n(x)$, is determined by the system configuration and is found by solving the Helmholtz equations.

**3. Lasing pre-threshold in toy-model of laser**

We begin our consideration with an analysis of a toy-model of a cavity-free system consisting of an ensemble of *N* two-level atoms located in a single point, $x = 0$, of the box. Recently, it has been shown that even a system having no cavity may lase.[37] We assume that the total number of two-level atoms is $5 \times 10^5$, the relaxation rates of atoms' polarization is $\gamma_\sigma = 10^{-2} \omega_{TLS}$, the relaxation rates of the population inversion is $\gamma_D = 10^{-6} \omega_{TLS}$, and $\gamma_a = 2 \times 10^{-3} \omega_{TLS}$, where $\omega_{TLS}$ is a transition frequency of two-level atoms.

The empty Universe is modeled as a large box. The modes of the box are standing waves with wavenumbers determined by the condition $k_n = 2\pi n / L_B$, where *n* is an integer. Also, we assume that the eigenfrequency of the *n*th mode is $\omega_n = c k_n$, where *c* is the speed of light. Since all the atoms are located at a single point, the field $\mathbf{E}_n(x_m)$ and the coupling constant $\Omega_{nm}$ are the same for all the atoms; and the latter is equal to $\Omega_0 = 3 \times 10^{-6} \omega_{TLS}$. This allows us to simplify Equation (1)-(3), by moving from the equations for polarizations and population inversions of each atom to the equations for the averaged over the atom ensemble values of these variables, $\sigma = \frac{1}{N} \sum_{m=1}^{N} \sigma_m$ and $D = \frac{1}{N} \sum_{m=1}^{N} D_m$.

Below the lasing threshold, the stationary values of the amplitudes, $a_n$, of modes of the EM field and the average values of atom polarizations, $\sigma$, are zero, while the average value of the atom population inversion is $D = (\gamma_P - \gamma_D)/(\gamma_P + \gamma_D)$.[33] To find the eigenfrequencies of small fluctuations of the amplitudes near the stationary state of the system, we linearize Equation (1)-(3) near the stationary state $a_n = \sigma = 0$ and $D = D_0 = (\gamma_P - \gamma_D)/(\gamma_P + \gamma_D)$. As a result, we obtain a closed system of linear differential equations for amplitudes of small fluctuations, $\delta a_n$, and the atom polarizations, $\delta \sigma$, which can be written in the matrix form:

$$\frac{d}{dt}\begin{pmatrix} \delta a_{-\infty} \\ ... \\ \delta a_{\infty} \\ \delta\sigma \end{pmatrix} = \begin{pmatrix} -\gamma_a/2 - i\Delta_{-\infty} & ... & 0 & -i\Omega_0 N \\ ... & ... & ... & ... \\ 0 & ... & -\gamma_a/2 - i\Delta_{\infty} & -i\Omega_0 N \\ i\Omega_0 D_0 & ... & i\Omega_0 D_0 & -\gamma_\sigma/2 \end{pmatrix} \begin{pmatrix} \delta a_{-\infty} \\ ... \\ \delta a_{\infty} \\ \delta\sigma \end{pmatrix}. \quad (4)$$

Now, we trace the dependencies of the eigenfrequencies of Equation (4) on the pump rate. We begin with studying the dynamics of a finite number of modes of the empty box, whose frequencies are in the range $(\omega_{TLS} - 3\gamma_\sigma, \omega_{TLS} + 3\gamma_\sigma)$. To find the eigenfrequencies of hybrid states, $\omega_j$, that include both the EM field and the polarization of atoms, we calculate the eigenvalues of the matrix in the right-hand part of Equation (4), $\lambda_j = -i\omega_j$. We use Implicitly Restarted Arnoldi Method (IRAM) from the Fortran library "ARPACK eigenvalues and eigenvectrors of large matrices" with double-precision complex arithmetic. This method is well suited for tackling large-scale problems.[38] The positions of the eigenfrequencies on the complex frequency plane for different pump rates are shown in **Figure 1**.

One can see that with an increase in the pump rate, the eigenfrequencies of the laser system move up in the complex plane (Figure 1). In addition, the real part of eigenfrequencies are pulled to the transition frequency of active atoms. There is a certain pump rate, at which two eigenfrequencies closest to the transition frequency coalesce (see also **Figure 2**). At this very point, a special state arises.

Above the pump rate corresponding to the coalescence point, the phenomenon of resonance trapping is observed.[10-15] Namely, if below the coalescence point, the imaginary parts of all eigenfrequencies move up toward the real axis, above this point, only the imaginary part of the frequency of the special state continues moving up. At the same time, the imaginary parts of all other eigenfrequencies move down away the real axis. Thus, with an increase in the pump rate, only the lifetime of the special state increases. Eventually, the eigenvalue of the special state reaches the real axis (see Figure 2b), and the special state starts lasing. This is the reason why we refer to the pump rate, at which the special state is formed, as a lasing pre-threshold.

The behavior of the eigenfrequencies in the complex frequencies plane near the lasing pre-threshold is characteristic to an exceptional point (EP).[8,9] It is known that the eigenstates, whose eigenfrequencies coalesce, are linearly dependent at the EP.[8,9] To show this, we consider the scalar product of the eigenstates. For this purpose, we present the eigenstates of the system as the vectors $\mathbf{e}_j = (\delta a_{-\infty}, ... \delta a_{\infty}, \delta\sigma)_j$. The scalar product between $j$th and $k$th eigenstates is determined as a scalar product of the corresponding vectors, i.e., $\mathbf{e}_j^T \cdot \mathbf{e}_k / \sqrt{\left(\mathbf{e}_j^T \cdot \mathbf{e}_j\right)\left(\mathbf{e}_k^T \cdot \mathbf{e}_k\right)}$. Such a scalar product is equal to zero only if the states are orthogonal, and it is equal to unity when the eigenstates coincide.

Below the lasing pre-threshold, an increase in the pump rate is accompanied by an increase in the scalar product of the eigenstates (see **Figure 3**). At the lasing pre-threshold, the two coalescent eigenstates coincide with each other (see Figure 3). That is, the scalar product between these eigenstates is equal to 1 up to a calculation error. Above the lasing pre-threshold, the scalar product between the eigenstates begins to decrease with an increase in the pump rate (Figure 3). Thus, in the toy-model, the lasing pre-threshold takes place in an exact EP, where the eigenstates, whose eigenfrequencies coalesce, become identical (see Figure 3).

Note that EPs are frequently observed in PT-symmetric lasers.[21,22,24-29,39] In such systems, the mode structure also changes in an EP and the long-living eigenmode forms.[21,22,24-29] In PT-symmetric systems, EPs appear due to the rebuilding of the EM field distribution between system regions with gain and loss.[31,40-46] This redistribution results in the appearance of eigenmodes localized in both system regions with gain and loss. The former eigenmodes are long-living, while the latter are short-living.

In conventional lasers with no particular symmetry, the mechanism of the formation of the special state is not connected with the rebuilding of the EM field distribution. In these lasers, an increase in the pump rate results in the change of the contribution of polarizations of atoms of the active medium in the hybrid eigenstates (see **Figure 4**). Below the lasing pre-threshold, the contribution of the atomic polarizations, $\sigma$, in all the eigenstates increases with an increase in the pump rate. Above the lasing pre-threshold, the contribution of the atomic polarization, $\sigma$, in the special state continues to grow with an increase in the pump rate (see the blue solid line in Figure 4). At the same time, the contribution of the atomic polarizations, $\sigma$, in all other eigenstates decreases (see Figure 4). An increase of the atomic polarization results in a boost of the interaction between the EM field of the modes and the inverted atoms. The energy flow from the atoms into a mode, which is proportional to $-ia^*\sigma$,[47] also increases. This increasing flow compensates for losses in the EM mode and leads to an increase in the imaginary part of the eigenfrequencies of the respective eigenstates (see Figure 2b).

Thus, it is the change of the atoms' polarization that causes the formation of the special state at a certain pump rate. In the toy-model of laser, the special state forms in the EP of the system, where two eigenstates coincide, and their eigenfrequencies are equal to each other.

When the pump rate increases further, the lasing begins at this special state. Therefore, the formation of the special state is the necessary condition of the lasing.

**4. Independence of the lasing pre-threshold on the box size**

In this section, we consider the effect of the box size, $L_B$, on eigenfrequencies of the system. Below the lasing pre-threshold, the eigenfrequencies of all eigenstates have similar dependencies on the box size. The coupling Rabi constant describing the interaction between a single EM mode and an atom, $\Omega_n = -\mathbf{E}_n(x=0) \cdot \mathbf{d}/\hbar$, is proportional to the amplitude of the quantum of the EM field in this mode. This amplitude is inversely proportional to $\sqrt{L_B}$.[36] For this reason, below the EP, the interaction between a single EM mode and an active atom decreases with an increase of $L_B$. Since the volume of the active medium remains unchanged, the interaction between a single EM mode and all atoms vanishes in the limit $L_B \to \infty$. The EM field distributions in the eigenstates are transformed into the distributions of the EM modes of the box without atoms. Therefore, in the complex plane, with an increase in $L_B$, the eigenfrequencies of all eigenstates move down toward the line $\omega = -i\gamma_a$. There are two eigenfrequencies that move slower than the rest. The closer the pump rate to the EP pump rate, $\gamma_{EP}$, the slower this movement. At the EP, the eigenfrequencies coalesce, and the special state arises. The lasing pre-threshold, $\gamma_{EP}(L_B)$ (i.e., the EP pump rate) and the eigenfrequency of the special state, weakly depend on the box size, reaching the finite values, $\gamma_{EP}(\infty)$, and $\omega_{sp}$ at $L_B = \infty$ (see **Figure 5**). The eigenfrequencies of the other eigenstates move down (see Figure 5b) reaching the line $\omega = -i\gamma_a$ at $L_B = \infty$.

Above the lasing pre-threshold, $\gamma_{EP}(L_B)$, the absolute value of the imaginary part of the eigenfrequency of the special state decreases. At the pump rate (the threshold), at which the imaginary part of this eigenfrequency becomes zero, the lasing begins (see the blue solid line in Figure 2b). Eigenfrequencies of both the lasing threshold and the special state weakly depend on $L_B$, approaching finite limits at $L_B = \infty$ (see Supporting Information for details).

**5. Lasing pre-threshold in the distributed cavity-free system**

In the previous section, we consider the formation of a special state in a toy-model of a laser, in which all active atoms are located at a single point. In a general case, when the active atoms occupy a region of a finite length, eigenfrequencies and eigenstates of the system are determined by the same Equation (1)-(3). The main difference is that the coupling constants, $\Omega_{nm}$, depend on atom positions, and the equations for the polarization and the population inversion of each atom cannot be reduced to equations for average values of these variables.

When the region occupied with active atoms has a finite length, there are interfaces between the region and environment. In this case, the laser is not cavity-free. The interfaces cause the reflectance so that the formation of the Fabry-Perot resonator. Then, at sufficiently high gain, lasing may occur.[48,49] However, the generation of coherent radiation can occur at the pump rate, which is several orders of magnitude smaller than that required for lasing due to the Fabry-Perot resonator.[37] Thus, at such pump rates, the system may be considered as a cavity-free. This situation is realized if we assume that two-level atoms are uniformly distributed within the region of the length $l = 10\lambda_{TLS}$, where $\lambda_{TLS} = 2\pi c / \omega_{TLS}$ is a wavelength at the transition frequency of atoms. The other parameters of the active atoms are the same as in the toy-model of laser.

The dependence of eigenfrequencies on the pump rate is shown in **Figure 6**. An EP, in which eigenfrequencies coalesce, arises only for a special set of parameters of the system. For an arbitrary length of the active layer, the eigenfrequencies do not coalesce (see Figure 6a). However, even if an EP is absent, its signature remains visible. The dependencies of the imaginary parts of the eigenfrequencies on the pump rate are similar to that for the toy-model (compare Figure 2b and 6b). At a certain pump rate, the special state is formed. At a further increase of the pump rate, the lasing begins at this special state. Thus, even in the absence of EP, the pre-threshold exists.

The exact value of the pump rate corresponding to the lasing pre-threshold can be defined as the pump rate, at which the eigenfrequencies of all eigenstates except the special state change the direction of their movement in the complex frequencies plane. The other way to define the lasing pre-threshold is based on the behavior of a scalar product of the special state and the neighboring state. The numerical simulation shows that the scalar product between the special state and the neighboring eigenstate achieves the maximum at a certain pump rate (**Figure 7**). When the pump rate increases further, the scalar product between the special state and all other eigenstates decrease. In the toy-model, the same behavior takes place after passing through the lasing pre-threshold, which coincides with the EP. For this reason, we can identify the lasing pre-threshold as the pump rate at which the scalar product reaches its maximum. Note that two definitions lead to the same value for the threshold pump rate.

A signature of the EP does not disappear with an increase in the box size, $L_B$. Regardless of the box size, the dependence of the imaginary part of an eigenfrequency of the special state demonstrates the threshold behavior. With an increase in the box size, eigenfrequencies of eigenstates move toward the line $\omega = -i\gamma_a$. However, above the lasing pre-threshold, there is one eigenstate - the special state - whose eigenfrequency does not depend on $L_B$. Thus, the signature of an EP plays a role similar to that of the EP discussed above. The pump rate at which an EP signature appears defines the lasing pre-threshold.

**6. Lasing pre-threshold in a distributed system with cavity**

In this section, we consider lasing of a system having a layer of an active medium placed in a Fabry-Perot cavity with semi-transparent mirrors. We assume that the layer of an active medium consists of $N$ two-level atoms uniformly distributed in the range from $-l/2$ to $l/2$. This region is inside the Fabry-Perot cavity of the length $l_{cav} = l$. The Fabry-Perot cavity is located in a uniform absorbing environment with the size $L_B \gg l_{cav}$.

To find eigenfrequencies and eigenstates of the system, we use linearized Equation (1)-(3). The coupling constants, $\Omega_{nm}$, depend on atom positions and on the spatial distribution of EM modes. Thus, the geometry of the optical system is specified by the coupling constants, $\Omega_{nm}$. We take into account a finite number of the EM modes of an empty box, whose frequencies are in the same range as considered above: $(\omega_{TLS} - 3\gamma_\sigma, \omega_{TLS} + 3\gamma_\sigma)$.

To begin with, we consider the case in which only one of the Fabry-Perot resonator modes lies within the linewidth of the active medium. We assume that $l_{cav} = l = 10\lambda_{TLS}$, the reflection coefficient of both mirrors forming the Fabry-Perot cavity is equal to $0.9i$; the parameters of the active medium are the same as in the previous section.

Positions of the eigenfrequencies in the complex frequency plane for different pump rates are shown in **Figure 8**a. Similar to the case of an extended cavity-free system, an EP may only exist at specific system parameters. However, the EP signature and the lasing pre-threshold, are always present. When the pump rate exceeds the pre-threshold, the imaginary part of the eigenfrequency of the special state rapidly increases (see Figure 8). When the imaginary part of the eigenfrequency of the special state turns to zero, in this eigenstate, the lasing begins. In addition, similar to the case of an extended cavity-free system, the scalar product of the special state and a neighboring state is maximal at the lasing pre-threshold (**Figure 9**). Thus, the process of the formation of the special state is similar to the one in the extended cavity-free system.

Now, we consider the case when several Fabry-Perot modes lie within the linewidth of the active medium. To do this, we increase the length of the active medium to $l_{cav} = l = 50.25 \lambda_{TLS}$; all other parameters remain the same. The length increase results in the appearance of a number of special states at corresponding lasing pre-thresholds. In particular, we observe two pre-thresholds (see **Figure 10**). Similar to the case of the EP signature in the single-mode lasers, the pre-thresholds appear at the pump rates, at which eigenfrequencies of all hybrid states except the special states change the direction of movement in the complex plane. Also, at the same pump rates, the scalar products of neighboring hybrid states have maxima. The movements of eigenfrequencies caused by changes in the box size, $L_B$, are similar to that in the previously considered systems. The final regime of the operation is determined by the mode competition mechanism.[34]

## 7. Consideration of experimental observation of the lasing pre-threshold

In previous sections, we predict the formation of special states at the lasing pre-thresholds in the various systems. The lasing pre-threshold can be detected by tracking the temporal dynamics of the EM field evolving from the stationary lasing regime after turning off the pump.

We consider a laser consisting of a layer of an active medium placed in a Fabry-Perot cavity with semi-transparent mirrors. We assume that the total number of two-level atoms is $6 \times 10^5$, the relaxation rates of the atoms' polarization, the population inversion, and the EM modes are $\gamma_\sigma = 10^{-2} \omega_{TLS}$, $\gamma_D = 10^{-6} \omega_{TLS}$, and $\gamma_a = 0.9 \times 10^{-4} \omega_{TLS}$, respectively, the coupling constant is $\Omega_0 = 4.7 \times 10^{-5} \omega_{TLS}$. We use Maxwell-Bloch Equations (1)-(3) for the simulation of the evolution of the laser intensity.

In the initial state of the system is set above the lasing threshold. After the stationary lasing regime has been established, the pumping should be turned off. After that, the EM field intensity, the atoms' polarization, and the population inversion of the active medium decrease over time. The relaxation rate of the population inversion is usually much lower than the relaxation rates of the EM field in the cavity and the polarization of active atoms (i.e., $\gamma_D \ll \gamma_\sigma, \gamma_a$).[33,34] For this reason, one can study a local variation of the EM mode amplitudes, $a_n$, and the atoms' polarization, $\sigma_m$, at some time $t_c$, neglecting the time variation of the population inversion. To do this, in the Tailor series for $D_m(t)$, we only retain the zeroth-order term $D_m^{(c)} = D_m(t_c)$. Then, in the vicinity of $t_c$, Equation (1) and (2) for the EM mode amplitudes, $a_n$, and the atoms' polarization, $\sigma_m$, become linear and can be written in the matrix form

$$\frac{d}{dt}\begin{pmatrix} a_{-\infty} \\ \ldots \\ a_{\infty} \\ \sigma_1 \\ \ldots \\ \sigma_N \end{pmatrix} = \begin{pmatrix} -\gamma_a/2 - i\Delta_{-\infty} & \ldots & 0 & -i\Omega_{-\infty 1} & \ldots & -i\Omega_{-\infty N} \\ \ldots & \ldots & \ldots & \ldots & \ldots & \ldots \\ 0 & \ldots & -\gamma_a/2 - i\Delta_{\infty} & -i\Omega_{\infty 1} & \ldots & -i\Omega_{\infty N} \\ i\Omega_{-\infty 1}D_1^{(c)} & \ldots & i\Omega_{-\infty N}D_N^{(c)} & -\gamma_\sigma/2 & \ldots & 0 \\ \ldots & \ldots & \ldots & \ldots & \ldots & \ldots \\ i\Omega_{\infty 1}D_1^{(c)} & \ldots & i\Omega_{\infty N}D_N^{(c)} & 0 & \ldots & -\gamma_\sigma/2 \end{pmatrix}\begin{pmatrix} a_{-\infty} \\ \ldots \\ a_{\infty} \\ \sigma_1 \\ \ldots \\ \sigma_N \end{pmatrix}. \quad (5)$$

Ultimately, we can introduce instantaneous eigenstates and eigenfrequencies, $\omega_j$, which are functions of $D_m^{(c)} = D_m(t_c)$. Note that the current values $D_m(t = t_c)$ are found by solving Maxwell-Bloch Equation (1)-(3).

Now we track the evolution of the instantaneous eigenstates and eigenfrequencies over time, starting with the stationary lasing state. This means that right after turning off the pumping, one of the instantaneous eigenfrequencies should be on the real axis (see **Figure 11**). We identify it as an eigenfrequency of the special state. The other eigenfrequencies lie near the line $\omega = -i\gamma_a$.

During the system relaxation, the imaginary part of the instantaneous eigenfrequency of the special state decreases (**Figure 12**a), and it rapidly moves fast down in the complex frequency plane. Simultaneously, the instantaneous eigenfrequencies of the other eigenstates move up (Figure 11). The motion of the eigenfrequencies allows one to determine the signature of the EP. At time, $t_{pre}$, the instantaneous eigenfrequencies of the special state and the neighboring eigenstate become close to each other (see Figure 11). Concurrently, the scalar product of these eigenstates takes a maximum value (Figure 12b). This is the time at which the signature of an EP and the lasing pre-threshold are observed. At $t > t_{pre}$ the imaginary parts of the eigenfrequencies of all eigenstates decrease (Figure 12a). Note that the rate of this decrease is significantly lower than the rate of the decrease for a special mode at $t < t_{pre}$.

Unfortunately, the motion of instantaneous eigenfrequencies $\omega_j$ cannot be measured directly in experiment. However, the change in the imaginary parts of the eigenfrequencies affects the relaxation rate of the EM field intensity, $I(t) = \sum_n |a_n(t)|^2$.

Solving Equation (1)-(3) we obtain the dependence $I(t)$ (**Figure 13**a). Immediately after the pumping is turned off, the EM field distribution in the laser coincides with that in the special state. During the system evolution, the EM field distribution changes only slightly, remaining almost the same as in the special state. The relaxation rate of the EM field intensity is determined by the imaginary part of the eigenfrequency of the special mode. A change of the imaginary part results in the change of the relaxation rate of the EM field intensity. As a result, the EM field intensity decays non-exponentially over time (Figure 13a) until the time when the pre-threshold is reached ($t < t_{pre}$). After crossing the lasing pre-threshold, the imaginary part of the eigenfrequency of the special mode almost stops changing over time (Figure 12a). The imaginary parts of all eigenfrequencies become practically the same and are equal to $\gamma_a$ (Figure 12a). As a result, the EM field energy begins decaying exponentially over time (**Figure 13**a). Thus, the transition from non-exponential to exponential decay can be interpreted as passing through the lasing pre-threshold. Note that the change in the EM field dynamics is not accompanied by a change in the dynamics of the population inversion, which slowly decreases over time (Figure 13b).

## 8. Conclusion

We demonstrate that in conventional laser systems without a particular symmetry, in addition to the ordinary lasing threshold, there is a lasing pre-threshold, at which a special state is born. At the lasing pre-threshold, the spectrum of the eigenstates changes drastically. Below the lasing pre-threshold, the lifetimes of all eigenstates increase with a pump rate. At the pre-threshold, two of the eigenmodes coalesce, forming an EP or coming close to each other (a signature of an EP). Above the lasing pre-threshold, the coalesced eigenmodes split again, and the lifetime of the special state continues to increase. At the same time, the lifetimes of all other eigenstates decrease. When the pump rate reaches the lasing threshold, lasing in the special state begins. The other eigenstates do not participate in the lasing. Thus, above the lasing pre-threshold, a laser transitions into a single-mode regime.

We demonstrate that the existence of the lasing pre-threshold (and the EP) is connected with the nature of the eigenstates of the laser system, which are hybrid states comprised of EM modes and atoms' polarization. Below the threshold, an increase in the pump rate only slightly increases the polarization contribution into eigenstates. Above the pre-threshold, the contribution of the atoms' polarization into the special eigenstate increases rapidly with an increase of the pump rate, while such a contribution into the other eigenstates decreases. As a result, in the special hybrid state, the coupling between the EM field and active atoms is intensified leading to lowering of the lasing threshold for this state.

We show that the lasing pre-threshold takes place in lasers with and without cavities. In the lasers, in which the cavity size is much larger than the generation wavelength, several special states are formed at their own lasing pre-thresholds. When the pump rate is greater than the lasing pre-thresholds, these special states behave like ordinary laser modes. In particular, with an increase in the pump rate, the frequencies of the special states are pulled to the transition frequency of the active medium.

Due to mode competition, the operation of a laser above the lasing threshold can often be considered as a single-mode system.[36,50] The obtained result shows that a decrease in the number of modes occurs above the pre-threshold. It is the competition of the special hybrid modes that results in the single-mode regime.

**Supporting Information**

Supporting Information is available from the Wiley Online Library or from the authors.

**Acknowledgements**

The work was supported by the Foundation for the Advancement of Theoretical Physics and Mathematics "BASIS." A.A.L. acknowledges the support of the ONR under Grant No. N00014-20-1-2198.

References

[1]  C. M. Bender, S. Boettcher, *Phys. Rev. Lett.* **1998**, *80*, 5243.
[2]  C. M. Bender, *Rep. Prog. Phys.* **2007**, *70*, 947.
[3]  R. El-Ganainy, K. G. Makris, M. Khajavikhan, Z. H. Musslimani, S. Rotter, D. N. Christodoulides, *Nat. Phys.* **2018**, *14*, 11.
[4]  S. Longhi, *Europhys. Lett.* **2018**, *120*, 64001.
[5]  M.-A. Miri, A. Alù, *Science* **2019**, *363*, eaar7709.
[6]  S. Özdemir, S. Rotter, F. Nori, L. Yang, *Nat. Mater.* **2019**, *18*, 783.
[7]  L. Feng, R. El-Ganainy, L. Ge, *Nature Photon.* **2017**, *11*, 752.
[8]  M. V. Berry, *Czech. J. Phys.* **2004**, *54*, 1039.

[9] N. Moiseyev, *Non-hermitian quantum mechanics,* Cambridge University Press, Cambridge, U, **2011**.

[10] P. Kleinwächter, I. Rotter, *Phys. Rev. C* **1985,** *32*, 1742.

[11] I. Rotter, *Rep. Prog. Phys.* **1991,** *54*, 635.

[12] E. Persson, I. Rotter, H. J. Stöckmann, M. Barth, *Phys. Rev. Lett.* **2000,** *85*, 2478.

[13] P. Seba, I. Rotter, M. Muller, E. Persson, K. Pichugin, *Phys. Rev. E* **2000,** *61*, 66.

[14] I. Rotter, E. Persson, K. Pichugin, P. Seba, *Phys. Rev. E* **2000,** *62*, 450.

[15] J. Okołowicz, M. Płoszajczak, I. Rotter, *Phys. Rep.* **2003,** *374*, 271.

[16] Y.-H. Lai, Y.-K. Lu, M.-G. Suh, Z. Yuan, K. Vahala, *Nature* **2019,** *576*, 65.

[17] H. Hodaei, A. U. Hassan, S. Wittek, H. Garcia-Gracia, R. El-Ganainy, D. N. Christodoulides, M. Khajavikhan, *Nature* **2017,** *548*, 187.

[18] W. Chen, S. K. Ozdemir, G. Zhao, J. Wiersig, L. Yang, *Nature* **2017,** *548*, 192.

[19] N. A. Mortensen, P. A. D. Goncalves, M. Khajavikhan, D. N. Christodoulides, C. Tserkezis, C. Wolff, *Optica* **2018,** *5*, 1342.

[20] Z. Dong, Z. Li, F. Yang, C. W. Qui, J. S. Ho, *Nat. Electron.* **2019,** *2*, 335.

[21] L. Feng, Z. J. Wong, R.-M. Ma, Y. Wang, X. Zhang, *Science* **2014,** *346*, 972.

[22] H. Hodaei, M.-A. Miri, M. Heinrich, D. N. Christodoulidies, M. Khajavikan, *Science* **2014,** *346*, 975.

[23] I. V. Doronin, A. A. Zyablovsky, E. S. Andrianov, A. A. Pukhov, A. P. Vinogradov, *Phys. Rev. A* **2019,** *100*, 021801(R).

[24] S. Longhi, *Phys. Rev. A* **2010,** *82*, 031801.

[25] Y. D. Chong, L. Ge, A. D. Stone, *Phys. Rev. Lett.* **2011,** *106*, 093902.

[26] B. Peng, Ş. K.Özdemir, S. Rotter, H. Yilmaz, M. Liertzer, F. Monifi, C. M. Bender, F. Nori, L. Yang, *Science* **2014,** *346*, 328.

[27] H. Hodaei, M. A. Miri, A. U. Hassan, W. E. Hayenga, M. Heinrich, D. N. Christodoulides, M. Khajavikhan, *Opt. Lett.* **2015,** *40*, 4955.

[28] M. Brandstetter, M. Liertzer, C. Deutsch, P. Klang, J. Schöberl, H. E. Türeci, G. Strasser, K. Unterrainer, S. Rotter, *Nat. Commun.* **2014,** *5*, 4034.

[29] M. Liertzer, L. Ge, A. Cerjan, A. D. Stone, H. E. Türeci, S. Rotter, *Phys. Rev. Lett.* **2012,** *108*, 173901.

[30] Z. Gu, N. Zhang, Q. Lyu, M. Li, S. Xiao, Q. Song, *Laser Photonics Rev.* **2016,** *10*, 588.

[31] K. G. Makris, R. El-Ganainy, D. N. Christodoulides, Z. H. Musslimani, *Phys. Rev. Lett.* **2008,** *100*, 103904.

[32] V. B. Berestetskii, E. M. Lifshitz, L. P. Pitaevskii, *Quantum electrodynamics,* Butterworth-Heinmann, **1982**.

[33] H. Haken, *Laser light dynamics,* North-Holland Physics Publishing, Oxford, UK, **1985**.

[34] A. E. Siegman, *Lasers,* University Science Books, Mill Valley, CA, **1986**.

[35] H. Carmichael, *An open systems approach to quantum optics,* Springer-Verlag, Berlin, **1991**.

[36] M. O. Scully, M. S. Zubairy, *Quantum optics,* Cambridge University Press, Cambridge, UK, **1997**.

[37] A. A. Zyablovsky, I. V. Doronin, E. S. Andrianov, A. A. Pukhov, Y. E. Lozovik, A. P. Vinogradov, A. A. Lisyansky, *Opt. Express* **2019,** *27*, 35376.

[38] R. B. Lehoucq, D. C. Sorensen, C. Yang, *Arpack users guide: Solution of large-scale eigenvalue problems with implicitly restarted arnoldi methods,* SIAM, Philadelphia, **1998**.

[39] Z. J. Wong, Y. L. Xu, J. Kim, K. O'Brien, Y. Wang, L. Feng, X. Zhang, *Nature Photon.* **2016,** *10*, 796.

[40] R. El-Ganainy, K. G. Makris, D. N. Christodoulides, Z. H. Musslimani, *Opt. Lett.* **2007,** *32*, 2632.

[41] S. Klaiman, U. Günther, N. Moiseyev, *Phys. Rev. Lett.* **2008,** *101*, 080402.

[42] A. Guo, G. J. Salamo, D. Duchesne, R. Morandotti, M. Volatier-Ravat, V. Aimez, G. A. Siviloglou, D. N. Christodoulidies, *Phys. Rev. Lett.* **2009,** *103*, 093902.

[43] C. E. Ruter, K. G. Makris, R. El-Ganainy, D. N. Christodoulides, M. Segev, D. Kip, *Nat. Phys.* **2010,** *6*, 192.

[44] A. Regensburger, C. Bersch, M. A. Miri, G. Onishcukov, D. N. Christodoulidies, U. Peschel, *Nature* **2012,** *488*, 167.

[45] B. Peng, Ş. K. Özdemir, F. Lei, F. Monifi, M. Gianfreda, G. L. Long, S. Fan, F. Nori, C. M. Bender, L. Yang, *Nature Phys.* **2014,** *10*, 394.

[46] A. A. Zyablovsky, A. P. Vinogradov, A. A. Pukhov, A. V. Dorofeenko, A. A. Lisyansky, *Phys. Usp.* **2014,** *57*, 1063.

[47] A. A. Zyablovsky, E. S. Andrianov, I. A. Nechepurenko, A. V. Dorofeenko, A. A. Pukhov, A. P. Vinogradov, *Phys. Rev. A* **2017,** *95*, 053835.

[48] A. V. Dorofeenko, A. A. Zyablovsky, A. A. Pukhov, A. A. Lisyansky, A. P. Vinogradov, *Phys. Usp.* **2012,** *55*, 1080.

[49] I. V. Doronin, E. S. Andrianov, A. A. Zyablovsky, A. A. Pukhov, Y. E. Lozovik, A. P. Vinogradov, A. A. Lisyansky, *Opt. Express* **2019,** *27*, 10991.

[50] R. Lang, M. O. Scully, W. E. Jr Lamb, *Phys. Rev. A* **1973**, *7*, 1788.

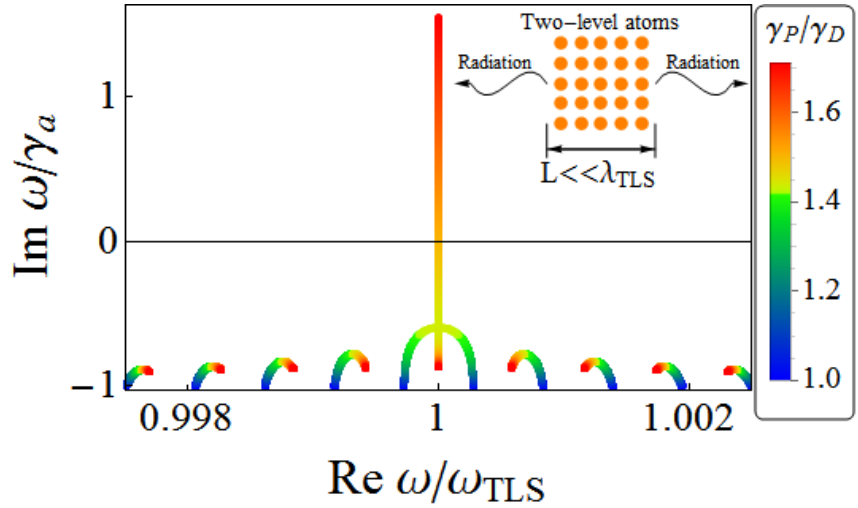

**Figure 1.** Trajectories of the eigenfrequencies in the complex frequency plane when the pump rate, $\gamma_P$, changes from $\gamma_D$ to $1.7\gamma_D$. The system size is $L_B = 1800\lambda_{TLS}$. Schematics of the system setup is shown in the inset.

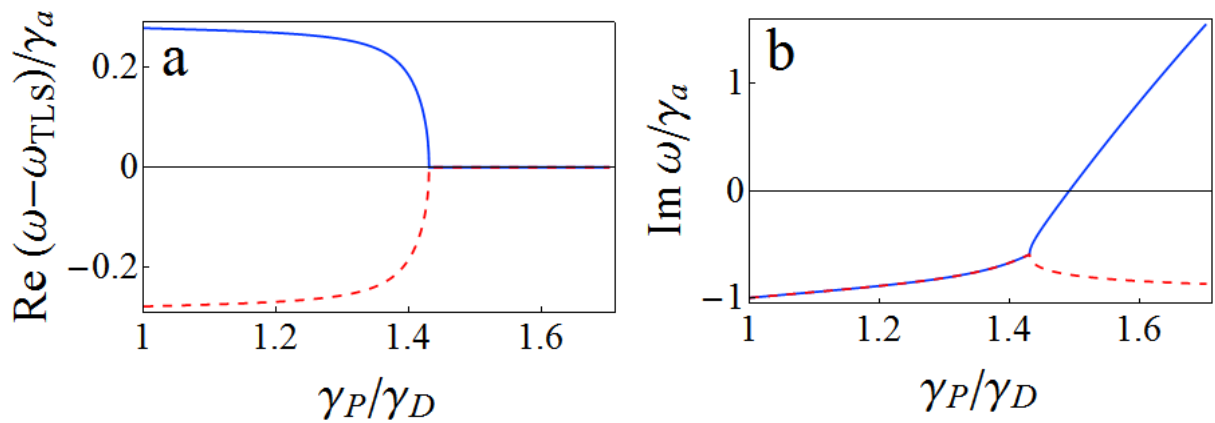

**Figure 2.** Dependences of the real (a) and imaginary (b) parts of the eigenfrequencies of the two coalescing eigenmodes on the pump rate. The modes are depicted by the blue solid and red dashed lines. The special state arises at $\gamma_P = 1.43\gamma_D$. The system size is $L_B = 1800\lambda_{TLS}$.

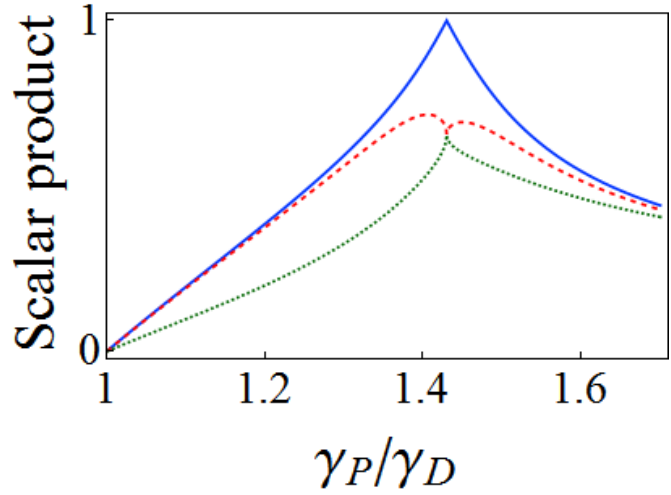

**Figure 3.** Pump rate dependence of scalar products of the mode with the lowest relaxation rate and the modes having the second (the blue solid line) and the third (the red dashed lines) lowest relaxation rates. (The former reaches unity in its maximum.) The same for the modes with the second and third lowest relaxation rates is shown by the green dotted line.

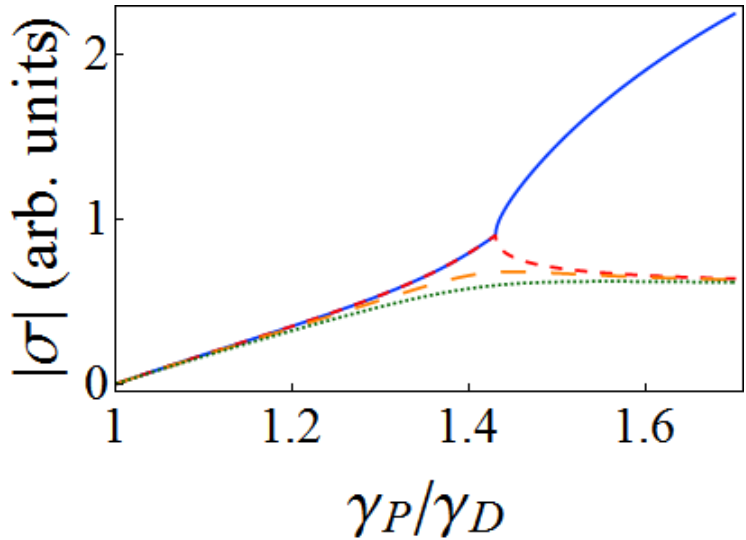

**Figure 4.** Dependence of the atomic polarization on the pump rate for four eigenmodes: two coalescent modes (shown by the blue solid and the red dashed lines) and their two neighbors (shown by the green dotted and orange dashed lines). The pump rate, at which the blue solid and red dashed lines split is the lasing pre-threshold.

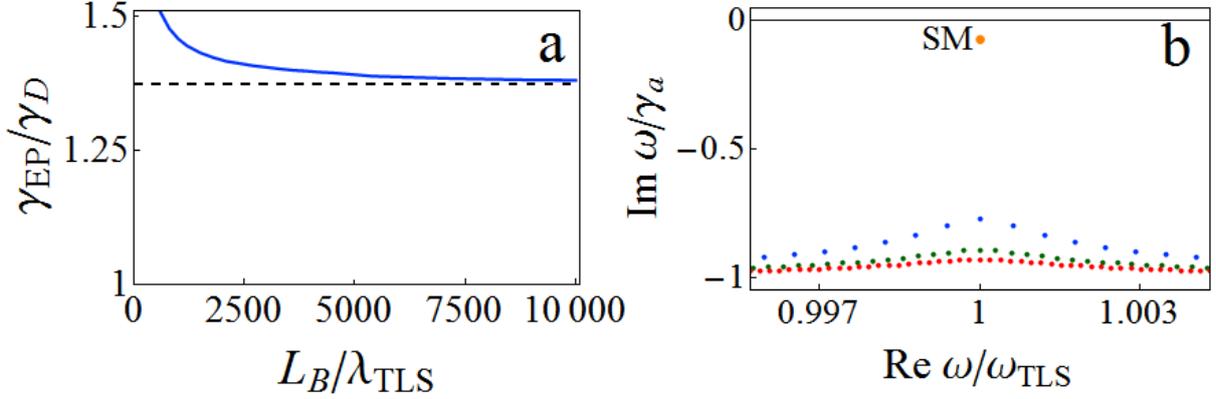

**Figure 5.** (a) Dependence of the pump rate (the lasing pre-threshold), at which the EP arises, on the box size, $L_B$. The dashed horizontal line shows the limit value of the lasing pre-threshold as $L_B \to \infty$ (see Supporting Information). (b) Positions of the eigenfrequencies above the pre-threshold for different box sizes: $L_B = 1800\,\lambda_{TLS}$ (blue dots), $L_B = 3600\,\lambda_{TLS}$ (green dots), and $L_B = 5400\,\lambda_{TLS}$ (red dots). The pump rate is $1.46\,\gamma_D$. The large orange dot corresponds to the special state. Its position does not depend on the box size.

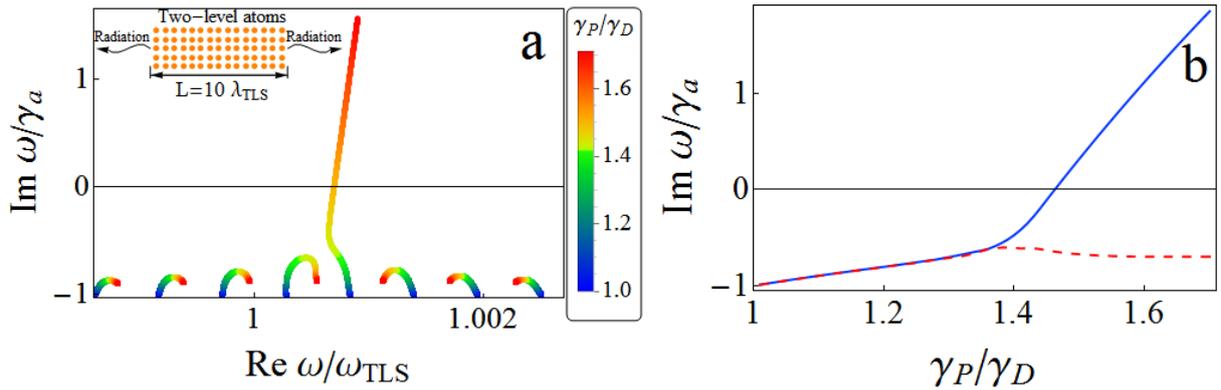

**Figure 6.** System with the distributed active medium showing the signature of the EP. (a) The trajectories of the eigenfrequencies in the complex frequency plane when the pump rate, $\gamma_P$, changes from $\gamma_D$ to $1.7\,\gamma_D$. (b) The dependencies of imaginary parts of the special state and the second eigenmode with the lowest relaxation rates on the pump rate. The length of the active layer is $l = 10\lambda_{TLS}$ and $L_B = 1800\lambda_{TLS}$. Schematics of the system setup is shown in the inset in Figure 6a.

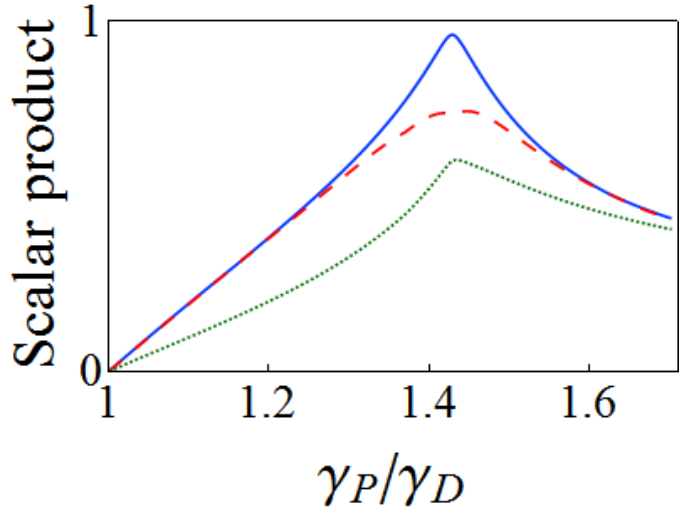

**Figure 7.** The same dependencies as in Figure 3 but for the system with active atoms distributed in a cavity. The maximum value of the scalar product is $0.9655$.

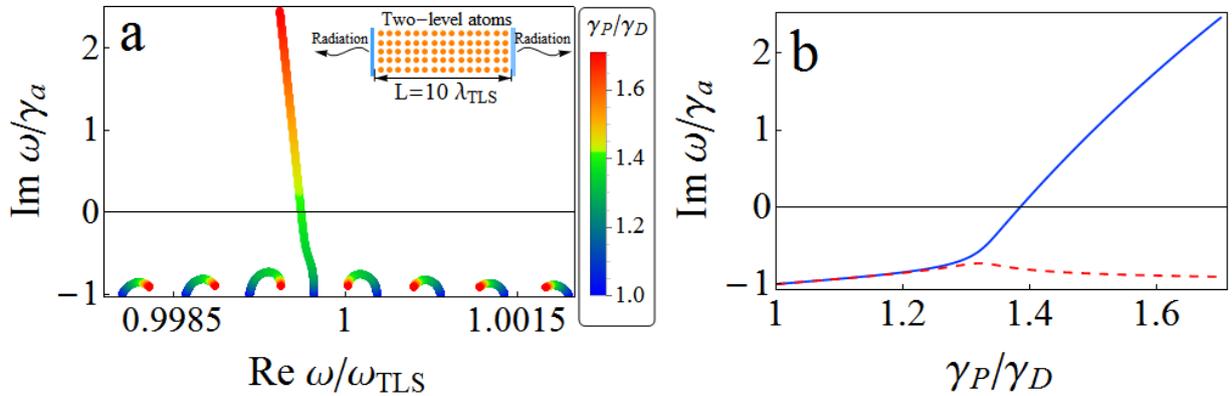

**Figure 8.** (a) Trajectories of the eigenfrequencies in the complex frequency plane when the pump rate, $\gamma_P$, changes from $\gamma_D$ to $1.7\gamma_D$. The lasing pre-threshold is $1.31\gamma_D$. (b) The dependencies of the imaginary parts of two eigenmodes with the lowest relaxation rates on the pump rate. The absolute value of the reflectance of the mirrors in the Fabry-Perot cavity is 0.9, the length of the active layer is $l = l_{cav} = 10\lambda_{TLS}$ and $L_B = 1800\lambda_{TLS}$. Schematics of the system setup is shown in the inset in Figure 8a.

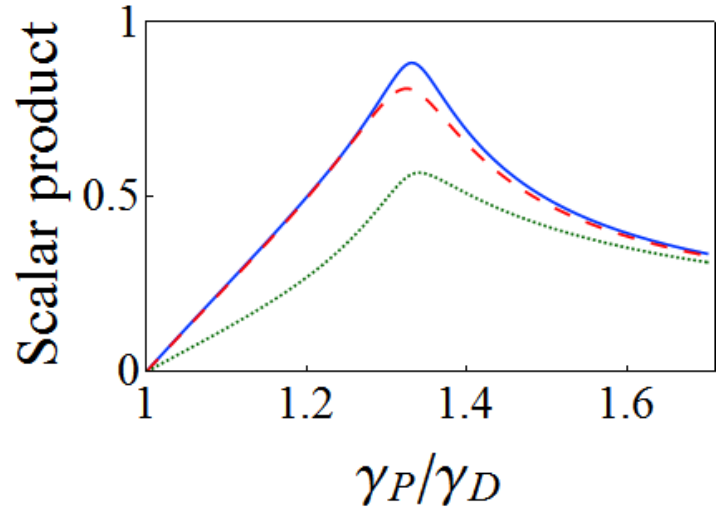

**Figure 9.** The same dependencies as in Figure 3 but for a system with a resonator The maximum value of the scalar product is $0.8842$.

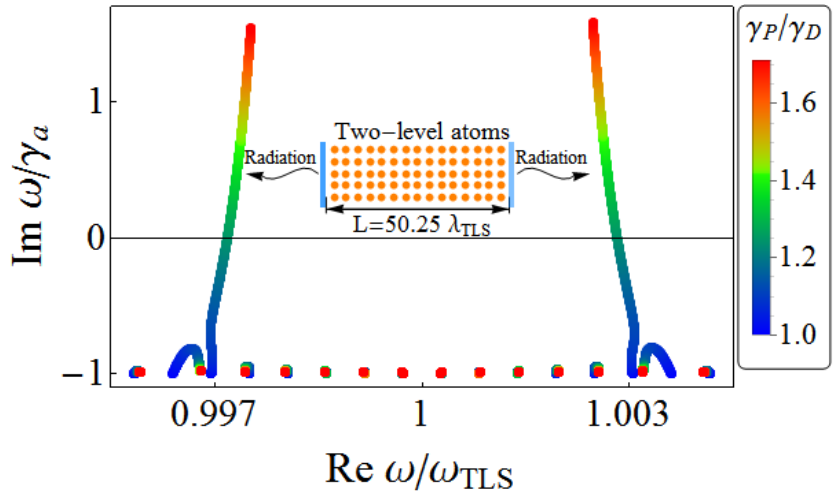

**Figure 10.** Trajectories of the eigenfrequencies in the complex frequency plane when the pump rate, $\gamma_P$, changes from $\gamma_D$ to $1.7\gamma_D$. The first lasing pre-threshold is $1.11\gamma_D$; the second lasing pre-threshold is $1.13\gamma_D$. The absolute value of the reflectance of the mirrors in the Fabry-Perot cavity is 0.9, the length of the active layer is $l = l_{cav} = 50.25\lambda_{TLS}$ and $L_B = 1800\lambda_{TLS}$. Schematics of the system setup is shown in the inset.

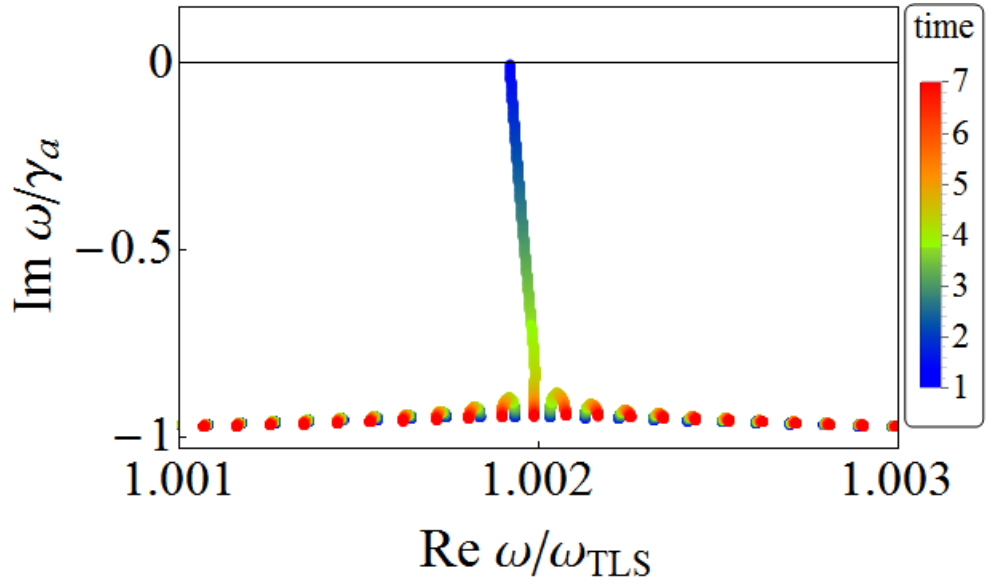

**Figure 11.** The trajectories of the eigenfrequencies in the complex frequency plane after the pumping is turned off.

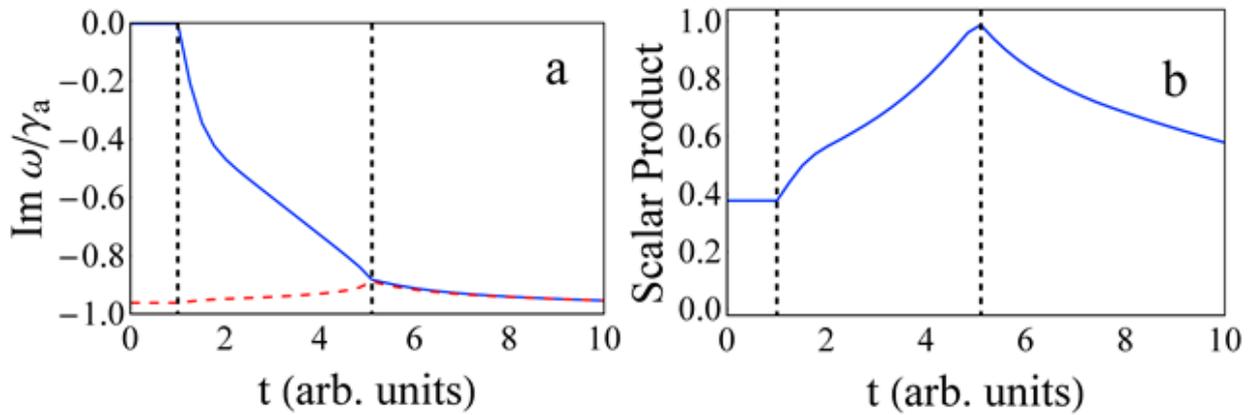

**Figure 12.** Time dependence of (a) the imaginary parts of instantaneous eigenfrequencies of two coalescent modes and (b) the scalar product of two coalescent modes.

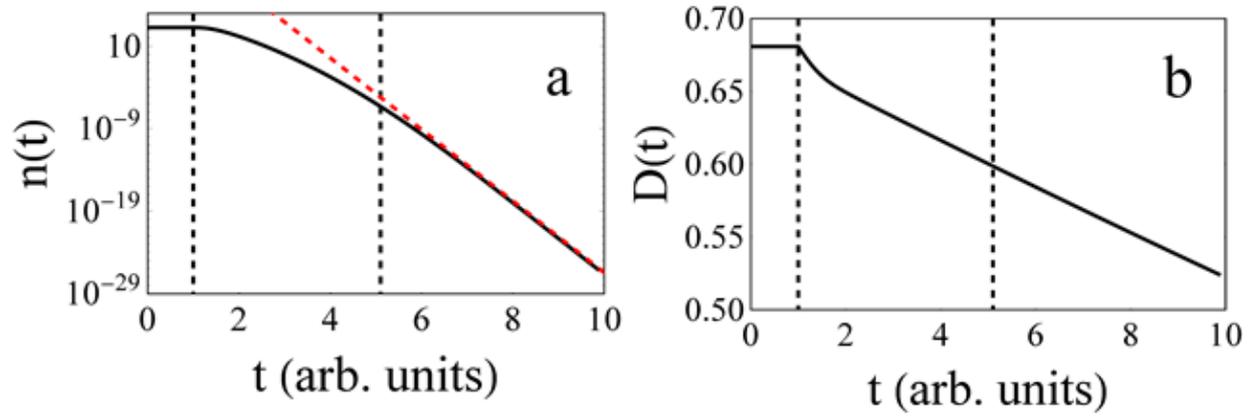

**Figure 13.** Time dependencies of (a) the EM field energy and (b) the population inversion. The left and right vertical black dashed lines show the time moments at which the laser crosses the lasing threshold and pre-threshold, respectively. The red dashed line shows the exponential decay with the decrement $\gamma_a$.

# Supporting Information

**Exceptional points as lasing pre-thresholds**

*Alexander A. Zyablovsky, Ilya V. Doronin, Eugeny S. Andrianov, Alexander A. Pukhov, Yurii E. Lozovik, Alexey P. Vinogradov, and Alexander A. Lisyansky\**

**Analytical expression for the lasing pre-threshold in a cavity-free system**

In the limit $L_B \to \infty$, we can obtain an analytical expression for the lasing pre-threshold in the toy-model of a cavity-free system, in which all active atoms are positioned at the same point. For this purpose, we find the eigenvalues of the matrix in the right part of Equation (4) of the main text, which are determined by the following equation

$$\left(\gamma_\sigma/2 + \lambda - \Omega_R^2 N D_0 \sum_{m=-\infty}^{\infty} \frac{1}{\gamma_a/2 + i\Delta_m + \lambda}\right) \prod_n (\gamma_a/2 + i\Delta_n + \lambda) = 0. \tag{S1}$$

In Equation (S1), the first factor becomes zero when

$$\lambda = -\gamma_\sigma/2 + \frac{2\Omega_R^2 N D_0}{\gamma_a/2 + \lambda} \sum_{m=-\infty}^{\infty} \frac{1}{1 + \left[\Delta(m-1/2)/(\gamma_a/2 + \lambda)\right]^2}. \tag{S2}$$

Since

$$\sum_{m=-\infty}^{\infty} \frac{1}{1 + (m-1/2)^2 a^2} = \frac{\pi}{2a} \tanh\left(\frac{\pi}{a}\right), \tag{S3}$$

we obtain that $\lambda$ satisfies the equation:

$$\lambda = -\gamma_\sigma/2 + \frac{\pi \Omega_R^2 N D_0}{\Delta} \tanh\left(\frac{\pi(\gamma_a/2 + \lambda)}{\Delta}\right). \tag{S4}$$

All parameters in Equation (S3) are real, and therefore, $\lambda$ is also real. Note that the real part of the eigenvalue is equal to the relaxation rate of the eigenmode, while its imaginary part is equal to the frequency detuning between the eigenfrequency and the transition frequency of the atom. Since, $\lambda$ is real, this frequency detuning is zero. Therefore, the eigenfrequency of the system is $\omega_{TLS} + i\lambda$.

When the box size tends to infinity, $\Delta = 2\pi c/L_B \to 0$. In this limit, the hyperbolic tangent turns into the step function, $\theta(\gamma_a/2 + \lambda)$, and Equation (S4) for the eigenvalue turns into

$$\lambda = \frac{\gamma_\sigma}{2}\left(-1 + \frac{D_0}{D_{th}^{(1)}} \theta(\gamma_a/2 + \lambda)\right), \tag{S5}$$

where $D_{th}^{(1)} = \gamma_\sigma \Delta / 2\pi \Omega_R^2 N$. When $D_0 = D_{th}^{(1)}$, the eigenvalue of the respective mode is zero that corresponds to the condition of the lasing threshold.[1]

For $0 < D_0 < D_{th}^{(1)}(\gamma_\sigma - \gamma_a)/\gamma_\sigma$ and $\gamma_\sigma > \gamma_a$ there is only one solution of Equation (S5)

$$\lambda_1 = -\frac{\gamma_\sigma}{2}\left(1 + \frac{D_0}{D_{th}^{(1)}}\right) < -\frac{\gamma_a}{2}. \tag{S6}$$

When $D_0 \geq D_{th}^{(1)}(\gamma_\sigma - \gamma_a)/\gamma_\sigma$, there is the second solution

$$\lambda_2 = \frac{\gamma_\sigma}{2}\left(-1 + \frac{D_0}{D_{th}^{(1)}}\right) \geq -\frac{\gamma_a}{2}. \tag{S7}$$

In the respective eigenmode, the energy flows from the atoms to the EM field ($\lambda_2 \geq -\gamma_a/2$), and the lasing can occurs at this mode.

The pump power at which this mode arises,

$$D_{th}^{(0)} = \frac{(\gamma_\sigma - \gamma_a)\Delta}{2\pi \Omega_R^2 N}, \tag{S8}$$

is the lasing pre-threshold in the cavity-free system considered. $D_{th}^{(1)}$ determines the lasing threshold for this mode.

Note that the lasing pre-threshold remains finite when $L_B \to \infty$. Indeed, the frequency difference between neighboring modes, $\Delta \sim 1/L_B$, while the coupling constant of the modes with the atoms, $\Omega_R \sim L_B^{-1/2}$. As a result, when $L_B \to \infty$, the lasing pre-threshold, Equation (S8), approaches a constant. A similar conclusion holds true for the lasing threshold, $D_0 = D_{th}^{(1)}$, which is determined by the condition $\lambda_2 = 0$.

Since we consider an absorbing environment, $\gamma_a > 0$, $D_{th}^{(1)}$ is larger than $D_{th}^{(0)}$, i.e., the lasing threshold is preceded by the lasing pre-threshold. This situation takes place, for example, in lasers used as sources in optical communication lines. The role of the absorbing environment is played by the optical communication lines, where there are always non-zero losses. If losses in the environment tend to zero, then the lasing pre-threshold and the lasing threshold coincide. This corresponds to a laser radiating into free space.

**References**

[1]    A. V. Dorofeenko, A. A. Zyablovsky, A. A. Pukhov, A. A. Lisyansky, and A. P. Vinogradov, *Phys. Usp.* **2012**, *55*, 1080.